\documentclass{emulateapj}

\usepackage{graphicx,graphics}
\usepackage{longtable}
\usepackage{amsmath}
\usepackage{amssymb}

\newcommand{\whathow}[1]{\noindent\textbf{What:} #1\\ \textbf{How/Why :} \noindent\begin{list}{$\bullet$}{\itemsep -1.3ex \topsep -1.3ex}}
\newcommand{\ewhathow}{\end{list}}
\newcommand{\figref}[1]{Figure~\ref{#1}}
\newcommand{\tabref}[1]{Table~\ref{#1}}
\newcommand{\eqnref}[1]{Equation~\ref{#1}}
\newcommand{\msun}{\ensuremath{M_\odot}}
\newcommand{\rsun}{\ensuremath{R_\odot}}
\newcommand{\logg}{\ensuremath{\log(g)}}
\newcommand{\target}{XMMU\,J013236.7+303228}
\newcommand{\lmc}{LMC\,X-4}
\newcommand{\smc}{SMC\,X-1}
\newcommand{\iraf}{\texttt{IRAF}}

\shorttitle{Constraints on the Compact Object Mass in the Eclipsing HMXB \target\ in M\,33}
\shortauthors{Bhalerao et al.}

\begin{document}
\title{Constraints on the Compact Object Mass in the Eclipsing HMXB \target\ in M\,33}

\author{Varun B. Bhalerao\altaffilmark{a}, Marten H van Kerkwijk\altaffilmark{b}, Fiona A. Harrison\altaffilmark{a}}

\altaffiltext{a}{Cahill Center for Astrophysics, California Institute of Technology, Pasadena, CA 91125, USA}
\altaffiltext{b}{Department of Astronomy and Astrophysics, 
University of Toronto, 50St. George Street, Toronto, ON M5S 3H4, 
Canada.}


\begin{abstract}
We present optical spectroscopic measurements of the eclipsing High Mass X-ray Binary \target\ in M\,33.  Based on
spectra taken at multiple epochs of the 1.73~d binary orbital period we determine physical as well as orbital
parameters for the donor star.   We find the donor to be a  B1.5IV sub-giant with effective temperature is $T=22,000 - 23,000$\,K.
From the luminosity, temperature and known distance to M33 we derive a radius of $R = 8.9\pm0.5\,R_\odot$.
    From the radial--velocity measurements, we determine a velocity semi-amplitude of
$K_{opt} = 63 \pm 12$\,km\,s$^{-1}$.    Using the physical properties of the  B-star determined from the
optical spectrum, we estimate the star's mass to be $M_{\rm opt} = 11\pm1$\,\msun .  Based on the X-ray spectrum,
the compact companion is likely a neutron star, although no pulsations have yet been detected.   Using the
spectroscopically derived B-star mass we find the neutron star companion mass to be  $M_{\rm X} = 2.0 \pm
0.4\,$\msun , consistent with the neutron star mass in the HMXB Vela X-1, but heavier than the canonical
value of 1.4~$\msun$ found for many millisecond pulsars.
We attempt to use as an additional constraint
that the B star radius inferred from temperature, flux, and distance,
should equate the Roche radius, since the system accretes by Roche lobe
overflow.  
This leads to substantially larger masses, but from trying to apply the
technique to known systems, we find that the masses are consistently
overestimated. Attempting to account for that in our uncertainties, we
derive $M_{\rm X}=2.2^{+0.8}_{-0.6}$\,\msun\ and $M_{\rm
opt}=13\pm4$\,\msun. We conclude that precise constraints require
detailed modeling of the shape of the Roche surface.
\end{abstract}


\maketitle

\section{Introduction}\label{sec:intro}

The range of possible neutron star masses depends on many factors, such as the initial mass of the progenitor's stellar core,  the details
of the explosion (in particular mass accretion as the explosion develops), subsequent mass accretion from a binary companion, and the pressure-density relation, or equation of state (EOS), of the neutron star matter.    
On the low-mass end, producing a neutron star requires the progenitor's core to exceed the 
Chandrasekhar mass, which depends on the uncertain electron fraction.  Theoretical models place this minimum 
mass in the range $M_{\rm core} \gtrsim 0.9$~--~1.3\,\msun\ \citep{tww96}.    The largest possible neutron star mass depends on the unknown physics
determining the EOS -- for example whether kaon condensates or strange matter can from in the interior \citep[See, for example,][]{lp05}.     The highest-mass neutron star to-date with an accurate measurement weighs in at $1.97 \pm 0.04~M_{\odot}$
\citep{dpr+10},  which already rules out the presence of  exotic hadronic matter at the nuclear saturation
density \citep{dpr+10,lp10}.

Testing the predictions of supernova models,  binary evolution models, and finding objects
at the extremes of the mass spectrum require determining neutron star masses in a variety of systems with differing 
progenitor masses and evolutionary history.
Neutron stars accompanying either a high-mass star or another neutron
star are thought
to have
accreted little to no matter over their lifetimes.  In contrast, neutron stars in low-mass X-ray binaries and millisecond
pulsars, typically in close orbits around a white dwarf, have undergone extended accretion periods that will make
the current mass exceed that at birth.   Different types of binaries will also have different average neutron star progenitor
masses.

High-mass X-ray Binaries (HMXBs) --  binaries containing a neutron star and a massive  ($\simeq20\,$M$_\odot$)
companion -- are particularly interesting systems in which to pursue mass measurements.  In most cases the neutron
star progenitor will have been more massive than the observed donor star,  yielding a relatively high-mass pre-supernova
core.    Furthermore, the NS mass will be close to the birth mass, since even for Eddington rates $\leq 0.1 M_{\odot}$
can be accreted in the $\sim 10^7$\,yr lifetime of the OB companion.   Indeed, among the five HMXB with reasonably 
secure masses, one (Vela X-1) has $M = 1.8$\,\msun\ \citep{bkv+01,qna+03}, indicating that this neutron star may have been born heavy.

Determining NS masses in HMXBs is, however, difficult.   In compact object binaries (NS -- NS, NS -- white
dwarf), highly precise mass measurements can be obtained from relativistic effects like the 
precession of periastron \citep{frb+08} or measurement of the Shapiro delay~\citep{dpr+10}.  In HXMBs,
however, accurate mass measurements are limited to eclipsing systems where orbital parameters for both
the NS and its stellar companion can be measured.   
For the NS this is done through X-ray or radio pulse timing,
and for the companion through radial--velocity measurements derived from doppler shifts in the stellar lines.
In the event pulsations are not detected, the NS mass can still be determined if good spectra are available to
estimate the mass of the optical component. In rare cases where the
distance to the binary is known, this provides an independent constraint
on the physical scale of the system -- for example by calculating the
absolute magnitude of the components. However, calculating masses from
such constraints is model-dependent.

\begin{figure}[htbp]
  \centering
  \includegraphics[scale=0.35]{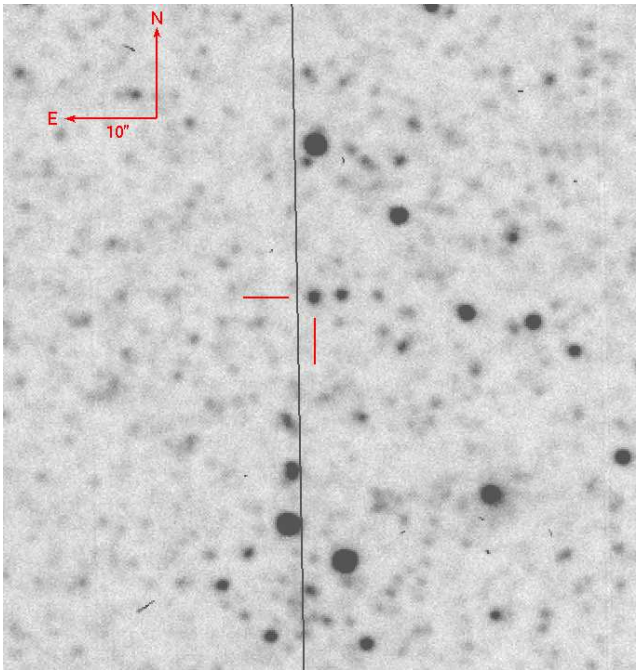}
  \caption{A V-band CFHT image showing the optical counterpart to \target, located at $\alpha = 01^{\rm h} 32^{\rm m} 36.^{\rm s}94$, $\delta= +30\degr 32\arcmin 28.\arcsec4$ (J2000). The image was obtained from CFHT online data archive, and a WCS was added using astrometry.net~\citep{lhm+10}.}
  \label{fig:finder}
\end{figure}

In this paper we present optical spectroscopic measurements of the donor star in the eclipsing HMXB \target\ using the Low Resolution Imaging Spectrograph on the 10\,m Keck-I telescope~\citep[LRIS;][]{occ+95} aimed at determining the mass of the compact companion.
\target\ was discovered by \citet{pmh+04} in their \emph{XMM-Newton} survey of \object[M 33]{M\,33}. In follow-up observations, \citet{pph+06,phg+09} identified it as an eclipsing High Mass X-ray Binary with a 1.73\,d period. The X-ray spectrum is hard, and the shape implies that 
the compact object is a neutron star. However no pulsations were detected in the X-ray data, so a black hole cannot be ruled out \citep{phg+09}. 
\citet{shm+06} discovered an optical counterpart (\figref{fig:finder}) which shows variability consistent with ellipsoidal modulation of the OB star. 
Given high quality spectra we are able to obtain a spectroscopic mass for the donor, and therefore determine
the compact object mass.  Using the known distance to M33 combined with the B stars luminosity and temperature
we derive a physical radius, which we equate with the Roche radius based on the observation that accretion is
occurring via Roche lobe overflow.    This provides an additional orbital constraint that we use to independently
estimate the compact object mass.

\section{Observations and data reduction}

We observed \target\ on UT 2009 October 16 and October 17, with the Low Resolution Imaging Spectrograph on the 10\,m Keck-I telescope~\citep[LRIS;][]{occ+95}, with upgraded blue~\citep{mcb+98,ssp+04} and red cameras~\citep{rsk+10}, covering a wavelength range from 3,200\,\AA{}\,--\,9,200\,\AA{}. We set up LRIS with the 600/4000 grism on the blue side and the 600/7500 grating on the red side, to get dispersions of 0.6\,\AA{}/pix and 0.8\,\AA{}/pix respectively (\tabref{tab:spectra}). To maximize stability of the spectra, we used the ``stationary rotator mode'', where the instrument rotator was held fixed near zero degrees rather than tracking the parallactic angle while observing. Atmospheric dispersion was compensated for by the ADC (Atmospheric Dispersion Corrector). We acquired a total of 28 spectra of the target, with exposure times ranging from 300--1800 seconds. The spectrophotometric standard EG\,247 was observed for flux calibration.

\begin{deluxetable*}{llcrr@{$\,\pm\,$}lcl}
\tabletypesize{\scriptsize}
\tablecaption{Details of individual exposures\label{tab:spectra}}
\tablewidth{0pt}
\tablehead{
\colhead{Image Name} & \colhead{MJD} & \colhead{Seeing\tablenotemark{a}} & \colhead{SNR} & \multicolumn{2}{c}{Heliocentric Velocity\tablenotemark{b}} & \colhead{$\chi^2$} &\colhead{Notes\tablenotemark{c}} \\
& & \colhead{arcsec} & &  \multicolumn{2}{c}{km\,s$^{-1}$} & &
}
\startdata
b091016\_0062 & 55120.276 & 1.0 &  5.2 & -106 &  24 & 0.96 & Rejected: Logs, Poor SNR \\
b091016\_0063 & 55120.283 & 1.0 & 21.7 &  -49 &   5 & 1.05 & Selected \\
b091016\_0064 & 55120.305 & 1.0 & 23.0 &  -32 &   5 & 1.08 & Selected \\
b091016\_0066 & 55120.333 & 1.0 & 22.8 &  -57 &   5 & 1.20 & Selected \\
b091016\_0080 & 55120.394 & 1.0 & 22.7 & -101 &   6 & 1.36 & Rejected: Logs \\
b091016\_0081 & 55120.418 & 1.0 &  8.7 & -109 &  15 & 0.99 & Rejected: Poor SNR \\
b091016\_0082 & 55120.425 & 1.1 & 20.1 &  -50 &   6 & 1.16 & Selected \\
b091016\_0087 & 55120.457 & 1.1 & 13.7 &  -47 &   9 & 1.09 & Selected \\
b091016\_0088 & 55120.474 & 1.4 &  4.2 &  -65 &  30 & 1.14 & Rejected: Logs, Poor SNR \\
b091016\_0098 & 55120.501 & 1.2 & 18.3 & -103 &   6 & 1.08 & Rejected: Logs \\
b091016\_0099 & 55120.524 & 1.1 & 19.0 &  -68 &   6 & 1.02 & Selected \\
b091016\_0106 & 55120.560 & 1.2 & 15.5 & -131 &   8 & 1.22 & Selected \\
b091016\_0108 & 55120.586 & 1.1 & 11.4 & -119 &  11 & 1.00 & Selected \\
b091016\_0109 & 55120.597 & 1.1 &  7.6 & -162 &  19 & 1.09 & Rejected: Poor SNR \\
b091016\_0110 & 55120.608 & 1.2 &  6.3 & -128 &  24 & 1.03 & Rejected: Logs, Poor SNR \\
b091017\_0057 & 55121.279 & 0.9 & 11.3 &  -42 &  11 & 0.99 & Rejected: Logs \\
b091017\_0059 & 55121.305 & 0.7 & 18.8 &  -96 &   6 & 0.94 & Selected \\
b091017\_0060 & 55121.327 & 0.9 & 19.6 &  -79 &   6 & 0.93 & Selected \\
b091017\_0068 & 55121.367 & 0.8 & 22.5 &  -75 &   5 & 0.93 & Selected \\
b091017\_0069 & 55121.389 & 0.8 & 22.5 &  -61 &   5 & 0.97 & Selected \\
b091017\_0071 & 55121.419 & 1.0 & 19.5 &  -77 &   6 & 0.99 & Selected \\
b091017\_0078 & 55121.457 & 1.0 & 19.7 &  -57 &   6 & 0.95 & Selected \\
b091017\_0079 & 55121.478 & 1.1 & 16.9 &  -27 &   7 & 1.03 & Selected \\
b091017\_0081 & 55121.507 & 1.1 & 18.0 &  -46 &   7 & 0.98 & Selected \\
b091017\_0087 & 55121.541 & 1.1 & 17.2 &  -44 &   7 & 1.02 & Selected \\
b091017\_0089 & 55121.567 & 1.1 & 15.1 &  -53 &   8 & 1.03 & Selected \\
b091017\_0090 & 55121.589 & 1.1 &  8.7 &  -70 &  13 & 0.89 & Rejected: Poor SNR \\
b091017\_0091 & 55121.600 & 1.1 &  8.4 &  -32 &  15 & 0.88 & Rejected: Poor SNR \\
\enddata
\tablecomments{\,LRIS was set up with a 1\arcsec\ slit, the D560 dichroic and clear filters on both red and blue arms. To maximize stability, we used the stationary rotator mode. The 600/4000 grism gives a dispersion of 0.61\,\AA{}/pix on the blue side. On the red side, we configured the 600/7500 grating at 27\degr.70 (central wavelength 7151\,\AA{}), and get dispersion of 0.80\,\AA{}/pix.}
\tablenotetext{a}{Seeing was measured as the median FWHM of Gaussians fitted to the trace of the blue side spectrum at several points.}
\tablenotetext{b}{Errors quoted here do not include an additional $15{\rm\,km\,s}^{-1}$ error to be added in quadrature, due to motion of star on the slit.}
\tablenotetext{c}{Spectra were not included in the final analysis if either the observing logs mentioned that the star had moved from the slit, or if the signal-to-noise ratio (SNR) per pixel in the extracted spectrum was below 10.}
\end{deluxetable*}

The data were reduced in \iraf\footnote{http://iraf.noao.edu/}. The spectra were trimmed and bias subtracted using overscan regions. No flatfielding was applied. Cosmic rays were rejected using \texttt{L.A.Cosmic} \citep{vandokkum01}. Atmospheric lines are stable to tens of meters per second~\citep{fpl+10}, so the wavelength solution for the red side was derived using sky lines for each image. For the blue side, the wavelength solution was derived from arcs taken at the start of the night. The spectra were then rectified and transformed to make the sky lines perpendicular to the trace, to ensure proper sky subtraction. The wavelength solutions for arcs taken at various points during the night are consistent with each other to a tenth of a pixel, with only an offset between different arcs. We corrected for this offset after extracting the spectra, by using the 5577.34\,\AA{} [O \texttt{I}] line. 
The spectra were extracted with \texttt{APALL}, and flux calibrated with data for EG\,247 and the standard \iraf\ lookup tables. We further tweaked the flux calibration by using a EG\,247 model spectrum from the HST Calibration Database Archive\footnote{ftp://ftp.stsci.edu/cdbs/current\_calspec/}. We used just one standard spectrum per night, and enabled airmass correction in \iraf\ during flux calibration.

\begin{figure*}[htbp]
  \centering
  \includegraphics[scale=0.7,angle=-90,viewport=130 230 550 500]{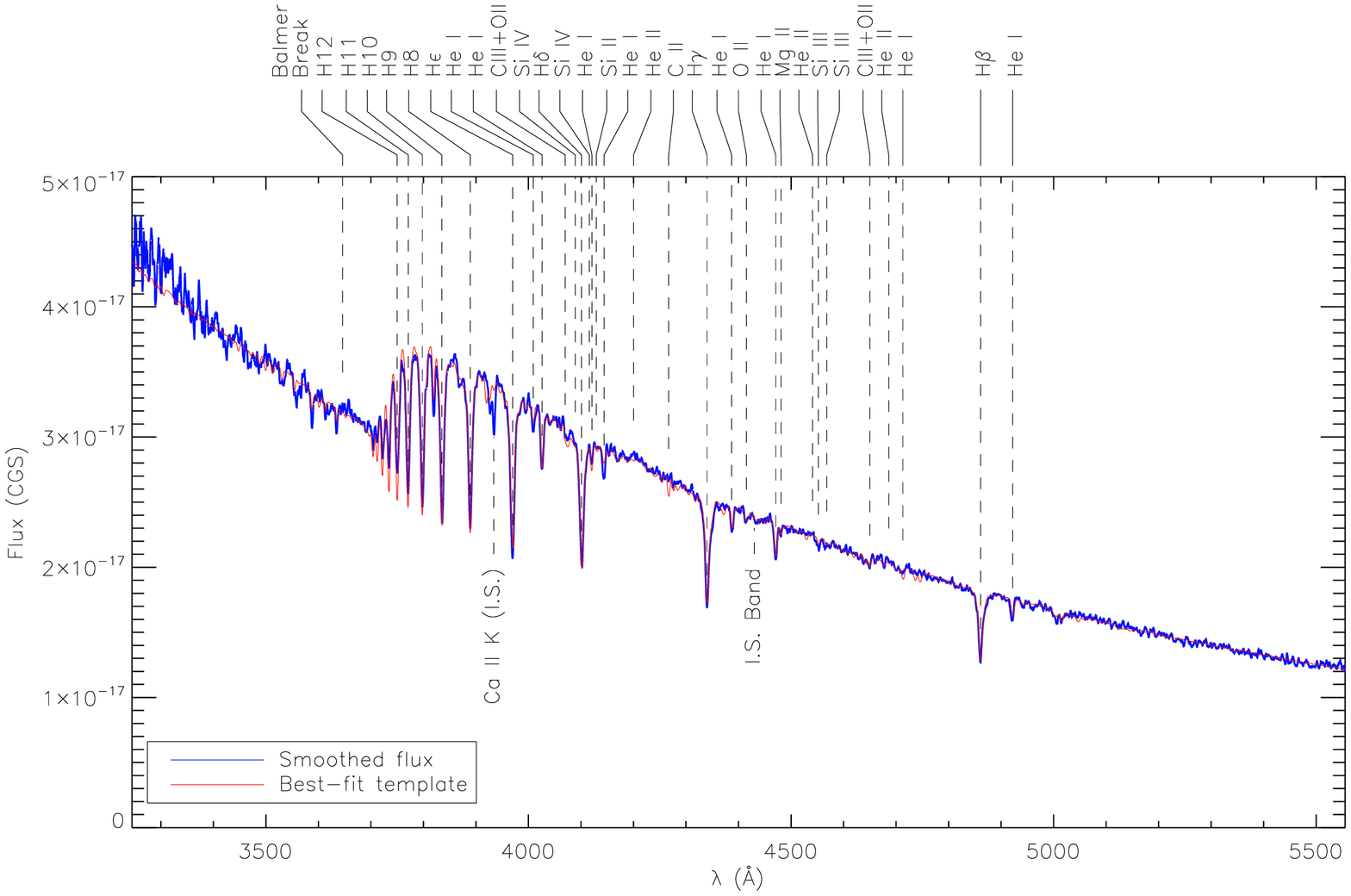}
  \caption{Observed spectrum and best-fit model for \target. The blue line is the average of the ten good spectra obtained on UT 2009 October 17, shifted to the rest wavelength using velocities from \tabref{tab:spectra}. The spectrum is smoothed with a 5 pixel (3\,\AA{}) boxcar for plotting. The red line is the best-fit template spectrum with $T=22000\,$K, $\log(g) = 3.5$, $v_{\rm rot}\sin i = 250{\rm\,km\,s}^{-1}$, solar metallicity. The template is reddened using $A_V = 0.395$, and scaled appropriately. Shifting the spectra to the rest frame blurs out the Ca \texttt{II} interstellar line and the 4430\,\AA{} interstellar band. The observed higher Balmer lines are less strong than those of the model, suggesting that the surface gravity is slightly higher than $\log g=3.5$ (consistent with our estimated spectral type and with our fit results; see text).}
  \label{fig:spectrum}
\end{figure*}

\begin{figure*}[htbp]
  \centering
  \includegraphics[scale=0.7,angle=-90,viewport=120 30 566 720]{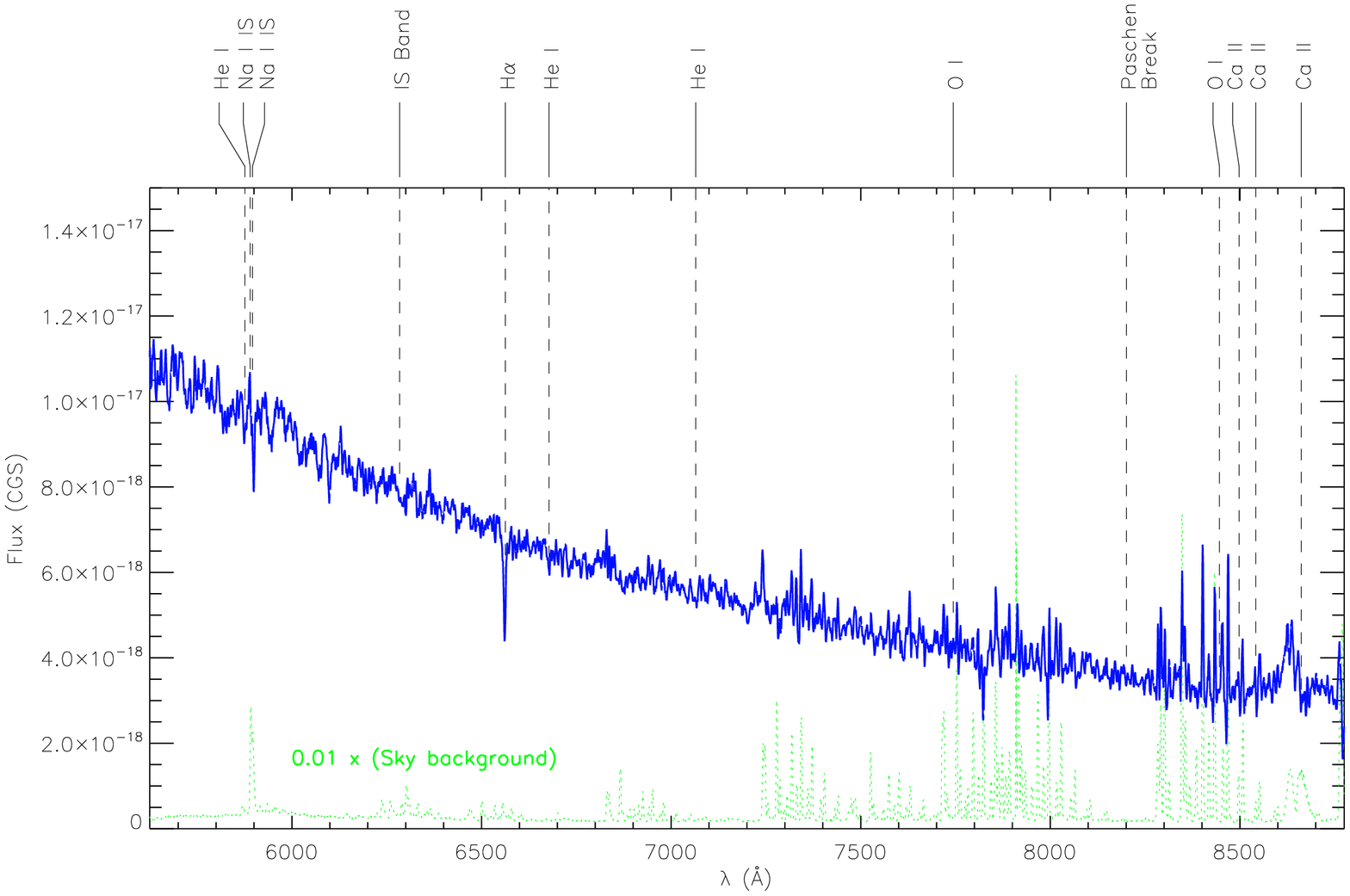}
  \caption{Observed red spectrum for \target. The blue line is the median combination of 48 spectra obtained on UT 2009 October 17, shifted to the rest wavelength using the velocity solution from \tabref{tab:params}. The flux axis differs from \figref{fig:spectrum}. The spectrum is smoothed with a 5 pixel (4\,\AA{}) boxcar for plotting. The source is much fainter than the sky in this wavelength region. For reference, the green dashed line shows the extracted sky spectrum, scaled \textit{down} by a factor of 100 for plotting. We do not use the red side spectra for any fitting.}
  \label{fig:redspectrum}
\end{figure*}

The final spectrum is shown in Figures \ref{fig:spectrum} \& \ref{fig:redspectrum}. The signal-to-noise ratio per pixel is $>10$ for most blue side spectra (\tabref{tab:spectra}). The first night, we experienced some tracking issues with the telescope, so the target did not remain well centered on the slit at all times. A similar problem was experienced for the first exposure on the next night, where the object was at a high airmass. In all following discussions, we reject some such spectra based on observing logs, and spectra with signal-to-noise ratio per pixel $<10$ (\tabref{tab:spectra}).

\section{Donor Star Parameters and Orbit}

We determine the best-fit stellar parameters and orbital solution using an iterative technique.  
First, we estimate a spectral type for the primary (donor) from individual spectra. We use appropriate spectral templates to calculate the orbital solution (Section \ref{subsec:orbit}). Next, we shift spectra to the rest frame and combine them to get a higher quality spectrum. We calculate stellar parameters from this combined spectrum, and use a template spectrum with these refined parameters to recalculate the velocities. In this section, we describe the final iterations of both these steps.

\subsection{Stellar parameters}\label{subsec:starpar}

Based on photometry of \target, \citet{phg+09} estimate that the companion is a 10.9\,\msun\ object with $T_{\rm eff} = 33000$\,K and $\log(g) = 4.5$, with $\chi^2 = 2.4$ for their best-fit model. They then assume a distance of 795\,kpc to M33, and calculate that the star has an absolute magnitude $M_V \sim -4.1$ and the line of sight extinction is $A_V = 0.6$, so derive a stellar radius of 8.0\,\rsun.

We deduce the spectral type by comparing our spectra to \citet{wf90} and the Gray spectral atlas\footnote{http://ned.ipac.caltech.edu/level5/Gray/frames.html}. The absence of He \texttt{II} lines (Figure~\ref{fig:spectrum}) implies a spectral type later than O, while the relative strengths of the Mg \texttt{II} 4482\,\AA{}/He \texttt{I} 4471\,\AA{} lines point to a spectral type earlier than B3. The strength of He \texttt{I} lines, and a weak feature at 4420\,\AA{} indicate a spectral type around B1, for main sequence stars. A bump blueward of H8 3889\,\AA{} is characteristic of spectral type B2. The weakness of C \texttt{III} 4650\,\AA{}, and the relative strengths of C \texttt{III}/O \texttt{II} near 4650\,\AA{} refine the spectral type to between B1 and B2 for both dwarfs and giants. Finally, the weakness of Si \texttt{IV} H$\delta$ 4101\,\AA{}, gives a spectral type of B1.5. To determine the luminosity class, we note that the O \texttt{II} 4415--4417\,\AA{} and Si \texttt{III} 4552\,\AA{} lines are present, but are weak as compared to He \texttt{I} 4387. We conclude that the donor is a B1.5IV star, with rough uncertainties of 0.5 spectral subclasses and one luminosity class.

Tabulated values of stellar parameters are usually provided for luminosity class III and V stars. For B1.5V stars, $T \approx 23000\,$K, $\log(g) = 4.14$, and $M_V = -2.8$. For B1.5III stars, $T \approx 22000\,$K, $\log(g) = 3.63$, and $M_V = -3.4$~\citep{cox00}. The B1.5IV target will have values intermediate to these. This inferred temperature is significantly cooler than $T \sim 33000\,$K reported by \citet{phg+09}. However, the absence of He II lines at, e.g., 4541 and 4686\,\AA, clearly exclude such a high temperature.
 

The colors of a star depend on its temperature and surface gravity. These expected colors can be compared to the observed colors to calculate the reddening and extinction. The expected color is $(B-V)_0 = -0.224$ for a 22000\,K sub-giant star, and $(B-V)_0 = -0.231$ for a 23000\,K main sequence star~\citep{bcp98}. From \citet{phg+09}, the mean magnitudes are $m_{g^\prime} = 21.03 \pm 0.02$, $m_{r^\prime} = 21.36 \pm0.02$. Using the \citet{jsr+05} photometric transformations for blue, $U-B<0$ stars\footnote{$V = g - 0.59(g-r) - 0.01 \pm 0.01$; $B-V = 0.90(g-r) + 0.21 \pm 0.03$.}, $m_V = 21.21 \pm 0.03$, $(B-V)_{\rm obs} = -0.09 \pm 0.04$. The color excess is $E(B-V) = (B-V)_{\rm obs} - (B-V)_0 = 0.14 \pm 0.04$. Using the standard ratio of total-to-selective extinction, $R_V = 3.1$ we get $A_V = 0.43 \pm 0.12$. For comparison, the foreground extinction to M33 is $A_V = 0.22$. 



We measure various stellar parameters by fitting our combined, flux calibrated spectrum with model atmospheres (taking into account the
 instrumental broadening; see Section~\ref{subsec:orbit}). For a Roche lobe filling companion (discussed in Section~\ref{sec:compmass}), we expect a radius of 6--10\,\rsun, surface gravity $\log(g) \simeq 3.7$ (consistent with our luminosity class), and projected rotational velocity $v_{\rm rot}\sin i \simeq 250{\rm\,km\,s^{-1}}$.
We use these values as starting points to select model atmospheres from a grid calculated by \citet{msc+05}. These templates are calculated in steps of 0.5\,dex in \logg, so we use models with $\log(g) = 3.5, 4.0$. For the initial fit, we assume $v_{\rm rot}\sin i = 250{\rm\,km\,s}^{-1}$ and solar metallicity. The only free parameters are a normalization and an extinction. We use extinction coefficients from \citet{cox00}, assuming $R_V = 3.1$. We find that the best-fit model for $\log(g) = 3.5$ has $T = 22100\pm40\,$K and $A_V = 0.401(3)$, with $\chi^2/{\rm DOF} = 1.13$ for 3600 degrees of freedom, while for $\log(g) = 4.0$, we get $T = 23500\pm50\,$K and $A_V = 0.425(3)$ with $\chi^2/{\rm DOF} = 1.22$. The temperatures are consistent with those expected for a B1.5 subgiant star, and the extinction is within the range derived from photometric measurements. \citet{phg+09} obtained a higher extinction for the target, which explains why they estimated the source temperature to be higher. \citet{msc+05} templates are calculated in temperature steps of 1000\,K in this range. For further analysis, we use the best-fit template: $\log(g) = 3.5$ and $T=22000\,$K. Since \logg\ is slightly higher than this value, for completeness we also give results using the best fit template for $\log(g) = 4.0$, which has $T = 23000\,$K. For both these templates, the best-fit extinction is $A_V = 0.395(3)$. We then keep $T$ and \logg\ constant and vary $v_{\rm rot}\sin i$. For both the \logg, $T$ combinations, we measure $v_{\rm rot}\sin i = 260\pm5{\rm\,km\,s}^{-1}$. Finally, using the same templates but with varying metallicity, we get the best fits for [M/H] = 0. The 0.5\,dex steps in [M/H] are too large to formally fit for uncertainties.

\begin{figure}[htbp]
  \centering
  \includegraphics[angle=-90,width=0.45\textwidth]{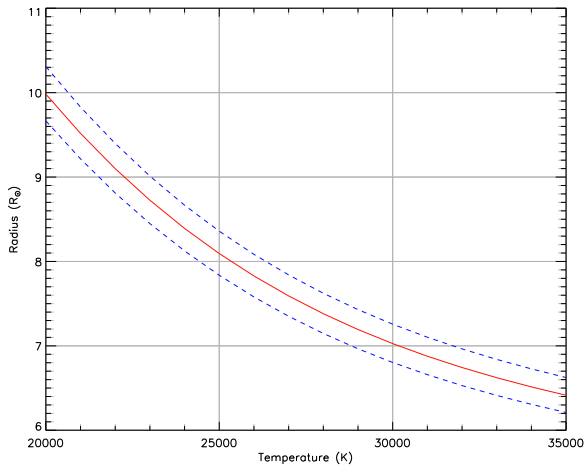}
  \caption{Radius -- temperature constraint from the observed luminosity. We calculate the absolute visual magnitude from the apparent magnitude, distance modulus and best-fit extinction. We then apply a temperature-dependent bolometric correction to calculate the bolometric magnitude. For any given temperature in this range, the uncertainty in radius is about 3\%.}
  \label{fig:rt}
\end{figure}

Next, we calculate the luminosity of the object to obtain a radius -- temperature relation. The distance modulus to M\,33 is $(m - M)_{\rm M33} = 24.54 \pm 0.06$ \citep[$d= 809$\,kpc;][]{mif+05,fmb+01}, which gives, accounting for the reddening of $A_V = 0.4$, $M_V = -3.74 \pm 0.07$. The bolometric luminosity of a star is related to its temperature and radius by $L_{\rm bol} \propto R^2 T^4$. To obtain the visual luminosity, one must apply a temperature-dependent bolometric correction, $BC = M_{bol}- M_V$. \citet{tor10} give formulae for bolometric correction as a power series in $\log (T)$. We calculate $BC = -2.11(-2.21)$ for $T=22000(23000)\,$K (which are consistent with \citet{bcp98} tables for main sequence stars.). After some basic algebra, we obtain:
\begin{eqnarray}\label{eq:rt}
&&  5 \log \left(\frac{R}{R_\odot}\right) + 10 \log \left(\frac{T}{T_\odot}\right) + BC(T) \\
&&  = M_{\rm bol,\odot} - m_V + (m-M)_{\rm M33} + A_V \\
&&  = 8.48 \pm 0.07 \nonumber
\end{eqnarray} 
Here, $M_{\rm bol,\odot} = 4.75$ \citep{bcp98}. The resultant radius-temperature relationship is shown in \figref{fig:rt}. For $T=22000(23000)\,$K, we infer $R = 9.1(8.7)\pm0.3\,R_\odot$. The absolute magnitude and radius are both consistent with a B1.5 sub-giant.

\subsection{Orbital parameters}\label{subsec:orbit}

We measure the radial velocities of the B star using model stellar spectra by \citet{msc+05}, as follows. For each observed spectrum, we measure the seeing using the width of the spectral trace. We then generate an instrument response function by taking a Gaussian matched to the seeing, truncating it at the slit size, and convolving it with the pixel size. Flux calibrated synthetic spectra (templates) are convolved with this instrument response, then redshifted to a test velocity. Then we redden the template using the measured value of extinction, $A_V = 0.395$, with coefficients from \citet{cox00}. We use \texttt{IDL}\footnote{http://www.ittvis.com/ProductServices/IDL.aspx} \texttt{mpfit}~\citep{markwardt09} to calculate the reddening and normalization to match this spectrum with the observed spectrum, and measure the $\chi^2$. By minimizing the $\chi^2$ over test velocities, we find the best-fit velocity and the error bars. This velocity is converted to a barycentric radial--velocity using the \texttt{baryvel} routine in Astrolib~\citep{landsman93}. \tabref{tab:spectra} lists the radial velocities for all spectra, measured using the best-fit stellar template: $T = 22000\,$K, $\log g = 3.5$, $[M/H] = 0.0$ and $v_{\rm rot}\sin i = 250{\rm\,km\,s}^{-1}$.

Red side spectra are not useful for radial--velocity measurement for several reasons. The flux of the B star at redder wavelengths is lower than the blue wavelengths, and there are fewer spectral lines in this range. Also, the background noise is higher, from the large number of cosmic rays detected by the LRIS red side and from intrinsic sky emission. Hence, all further discussion omits red side spectra.

We calculate an orbital solution for the B star using these radial velocities. Owing to the short 1.73\,d period of the system, we assume that the orbit must be circularized. The orbital solution is then given by:
\begin{equation}\label{eq:circorbit}
  v(t) = \gamma_{\rm opt} + K_{\rm opt} \sin \left(2 \pi \frac{t - T_0}{P} \right)
\end{equation} 
where $\gamma_{\rm opt}$ is the systemic velocity, $K_{\rm opt}$ is the projected semi-amplitude of radial--velocity, and $T_0$ is the epoch of mid-eclipse. We adopt $P = 1.732479 \pm 0.000027$ and $T_0 = 2453997.476 \pm 0.006$ from \citet{phg+09}. We obtain $\gamma_{\rm opt} = -80 \pm 5{\rm\,km\,s}^{-1}$, and $K_{\rm opt} = 64 \pm 12{\rm\,km\,s}^{-1}$. 
The best-fit has $\chi^2$/DOF = 5.1 for 16 degrees of freedom, which is
rather poor.  We find that an additional error term $\Delta v = 15{\rm\,km\,s}^{-1}$ needs to be added in quadrature to our error estimates to obtain $\chi^2_{\rm red} = 1$. We attribute this to movement of the star on the slit. If the target has systematic offset of 0\arcsec.1 from the slit center over the entire 30\,min exposure, it shifts the line centroids by about 0.45 pixels or 18\,km\,s$^{-1}$. For comparison, \citet{vbk11} find a similar scatter (13\,km/s$^{-1}$) in their observations of a reference star when using LRIS with a similar configuration (600/4000 grating, 0\arcsec .7 slit). Future observations should orient the spectrograph slit to obtain a reference star spectrum to correct for such an offset.

\begin{figure}[htbp]
  \centering
  \includegraphics[angle=-90,viewport=18 6 286 438,width=0.45\textwidth]{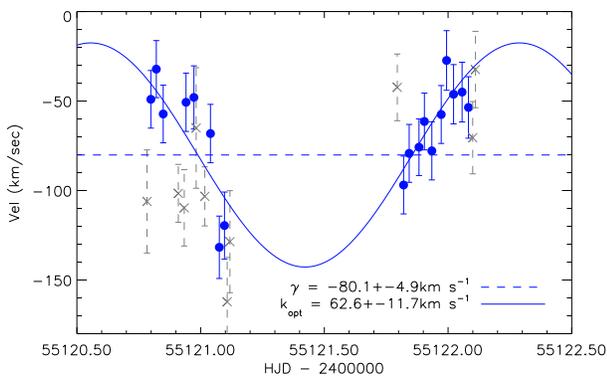} 
  \caption{The radial--velocity curve for \target. Solid blue circles denote points which were used in the final fit, and gray crosses are points which were rejected based on observing logs or a signal-to-noise ratio under 10 (\tabref{tab:spectra}). The solid blue curve is the best-fit orbital solution, and the dashed blue line is the mean systemic velocity.}
  \label{fig:velocity}
\end{figure}

\begin{deluxetable}{lr@{\,}l}
\tablecaption{System parameters for \target\label{tab:params}}
\tablewidth{0pt}
\tablehead{
\colhead{Property} & \multicolumn{2}{c}{Value}
}
\startdata
\sidehead{From \citet{phg+09}}
Period ($P$) & 1.732479 & $\pm 0.000027$\\
HJD of mid-eclipse ($T_0$) & 2453997.476 & $\pm 0.006$\\
Eclipse half-angle ($\theta_e$) & 30\degr.6 & $\pm 1\degr.2$ \\
\sidehead{This work}
Systemic velocity\tablenotemark{a} ($\gamma_{\rm opt}$) & $-80$ & $\pm 5{\rm\,km\,s}^{-1}$\\
Velocity semi-amplitude\tablenotemark{a} ($k_{\rm opt}$) & 63 & $\pm 12{\rm\,km\,s}^{-1}$\\
HJD of mid-eclipse\tablenotemark{b} ($T_0$) & 2453997.489 & $\pm 0.019$\\
\sidehead{Spectroscopically inferred}
OB star spectroscopic mass ($M_{\rm opt}$) & $11$ & $\pm 1$\,\msun\\
NS mass ($M_{\rm X}$) & 2.0 & $\pm0.4$\,\msun\\
\sidehead{Distance-based calculations}
OB star mass ($M_{\rm opt}$) & 18.1 & $^{+2.0}_{-1.9}$\,\msun\\
NS mass ($M_{\rm X}$) & 2.7 & $^{+0.7}_{-0.6}$\,\msun\\
\enddata
\tablenotetext{a}{Calculated using $P = 1.732479 \pm 0.000027$ and $T_0 = 2453997.476 \pm 0.006$ from \citet{phg+09}.}
\tablenotetext{b}{When we allow $T_0$ to vary, we obtain $\gamma_{\rm opt} = -80 \pm  5{\rm\,km\,s}^{-1}$ and $K_{\rm opt} = 64 \pm 12{\rm\,km\,s}^{-1}$. The analysis in this paper uses the \citet{phg+09} value of $T_0$.}
\end{deluxetable}

When we fit Equation~\ref{eq:circorbit} to data including the $15{\rm\,km\,s}^{-1}$ error in quadrature, we obtain $\gamma_{\rm opt} = -80 \pm 5{\rm\,km\,s}^{-1}$, and $K_{\rm opt} = 63 \pm 12{\rm\,km\,s}^{-1}$ (\figref{fig:velocity}, \tabref{tab:params}). 
For the epoch of observations, the uncertainty in phase is 0.019\,d. If we allow $T_0$ to vary, we get $T_{0,{\rm fit}} = 2453997.489\pm0.019$ (Heliocentric Julian Date), $\gamma_{\rm opt} = -80 \pm  5{\rm\,km\,s}^{-1}$, and $K_{\rm opt} = 64 \pm 12{\rm\,km\,s}^{-1}$. These values are consistent with those obtained using the \citet{phg+09} ephemeris. Hence, for the rest of this paper, we simply assume their best-fit value for $T_0$.

To investigate the sensitivity of the result to the choice of the stellar template, we repeat the measurement with a variety of templates. We vary the temperature from O9 (33,000\,K) to B3 (18,000\,K) spectral classes. As before, we use templates with $\log(g) = 3.5, 4.0$; $v_{\rm rot}\sin i = 250{\rm\,km\,s}^{-1}$ and solar metallicity. Repeating the orbit calculations for each of these models, we find that the systemic velocity $\gamma_{\rm opt}$ may change between models being fit: the extreme values are $-77 \pm 6{\rm\,km\,s}^{-1}$ and $-90 \pm 5{\rm\,km\,s}^{-1}$, a $1.7 \sigma$ difference. We suspect that the reason for this variation is the difference in shape of the continuum, as the extinction was held constant in these fits. For hotter templates with a steeper continuum, the red side of a line has lower flux than the blue side, so lowest $\chi^2$ will be obtained at a slightly higher redshift, as seen. The magnitude of this effect should be independent of the intrinsic Doppler shift of the spectrum, and should not affect the velocity semi-amplitude $K_{\rm opt}$. This is indeed the case: $K_{\rm opt}$ is constant irrespective of templates. The extreme values are $62 \pm 12{\rm\,km\,s}^{-1}$ and $63 \pm 12{\rm\,km\,s}^{-1}$, differing by less than $0.1 \sigma$. Dynamical calculations depend only on $K_{\rm opt}$, hence they are robust to the selection of template.

\section{Component masses}\label{sec:compmass}

The general method for accurately determining masses 
in X-ray binaries \citep{jr84} requires measuring the orbit
for both components, as well as having a constraint on the orbit inclination.  In general, the mass ($M_1$) of a component is expressed in terms of five parameters: the orbital period $P$, the radial--velocity semi-amplitude of the companion ($K_2$), the eccentricity $e$, the orbital inclination $i$ and mass ratio $q=M_1/M_2$. The first three parameters can be readily obtained by characterizing the orbit of either component through pulse timing of the NS in the X-ray or radio, or by spectroscopically measuring the radial--velocity of the optical companion at optical or infra-red wavelengths. Determining the mass ratio requires measuring orbital parameters for both components: $q = K_2/K_1$.   
In an eclipsing system, the inclination can be
constrained to be nearly edge-on ($i \approx 90\degr$), with a lower limit derived from eclipse duration and Roche-lobe arguments. Using these measurements, masses of both components in the system can be directly determined 
without model assumptions~\citep[see for example,][]{vkv+07,mnc+11}. 

Because no pulsations have been detected from the compact object in \target\ \citep{phg+09}, we need one more constraint in 
addition to the radial--velocity semi-amplitude of the donor. In Section~\ref{subsec:specmass} we use the spectroscopically inferred mass of the primary to calculate mass of the secondary from the mass ratio.    Because we know the distance to M33, and hence to \target, we can
also estimate the physical size of the secondary from the distance, its luminosity and temperature.   This provides a cross check
on the mass determination derived from the spectroscopic donor mass (Section~\ref{subsec:rochelims}).

\subsection{The spectroscopic method}\label{subsec:specmass}

In the following, we will denote the masses of the compact object and  donor star by $M_{\rm X}$ and $M_{\rm opt}$ respectively.  
The mass of the compact object ($M_{\rm X}$) is related to the radial--velocity of the B star ($K_{\rm opt}$) as follows:
\begin{equation}\label{eq:mx}
M_{\mathrm{X}} = \frac{K_{\mathrm{opt}}^3 P(1 - e^2)^{3/2}}{2\pi G \sin^3 i} \left(1+\frac{1}{q}\right)^2
\end{equation}
where $q = M_{\rm X}/M_{\rm opt}$ is the ratio of masses, defined so that higher values of $M_{\rm X}$ relate to higher values of $q$. $P$ is the orbital period of the binary, $e$ is the eccentricity, and $i$ is the inclination of the orbit. For eclipsing systems, the inclination is constrained by,
\begin{equation}\label{eq:sinithetae}
\sin i = \frac{\sqrt{1 - \beta^2\left({R_{\rm L}}/{a} \right)^2}}{\cos \theta_{\rm e}}
\end{equation}
where $R_{\rm L}$ is the volume radius of the Roche lobe, $a$ is the semi-major axis, and $\beta$ is the Roche lobe filling factor \citep{jr84}. For \target, the eclipse half-angle is $\theta_{\rm e} = 30\degr.6 \pm 1\degr.2$ \citep{phg+09}. Owing to the short orbital period, we assume that the orbit is circular and the B star rotation is completely synchronized with its orbit. For co-rotating stars, \citet{eggleton83} expresses $R_{\rm L}/a$ in terms of $q$:
\begin{equation}\label{eq:rla}
  \frac{R_{\rm L}}{a} = \frac{0.49 q^{-2/3}}{0.6q^{-2/3} + \ln (1 + q^{-1/3})}
\end{equation} 

The constant, relatively high X-ray luminosity, sustained over the non-eclipsed parts of the orbit, strongly indicates that accretion is occurring via
Roche lobe overflow.   In Roche lobe overflow, matter flowing through the Lagrangian point may form a disc around the compact object before being accreted onto it. This disc can occult the compact object, causing periods of low X-ray luminosity.  Both these characteristics are seen in the X-ray lightcurves of \target~\citep{phg+09}.  If mass is being accreted onto an object in a spherically symmetric manner, the accretion rate is limited by the Eddington rate, $\dot{M}_{\rm Edd}$ and the peak luminosity is $L_{\rm Edd}/L_\odot = 3\times 10^4 M/M_\odot$. For a 1.5\,--\,2.5\,\msun\ compact object, $L_{\rm Edd} = 1.8 - 3 \times 10^{38}{\rm\,erg\,s}^{-1}$. At its brightest, the source luminosity in the 0.2\,--\,4.5\,keV band was $2.0\times10^{37}{\rm\,erg\,s}^{-1}$ --- about $0.1 L_{\rm Edd}$. This luminosity was sustained throughout {\rm Chandra} ObsID 6387, which covered about 0.72\,d of the non-eclipsed orbit~\citep{phg+09}. Comparable flux was observed in the non-eclipsed parts of the orbit (0.73\,d) in {\rm Chandra} ObsID 6385.  Such high luminosity sustained over significant parts of the orbit is not observed in wind-fed systems, which have
typical luminosities an order of magnitude smaller.  Further, the short 1.73~d orbital period is not consistent with Be X-ray binary or wind-fed systems.  We conclude therefore the B star fills its Roche lobe.

\begin{figure}[htbp]
  \centering
  \includegraphics[angle=-90,width=0.5\textwidth]{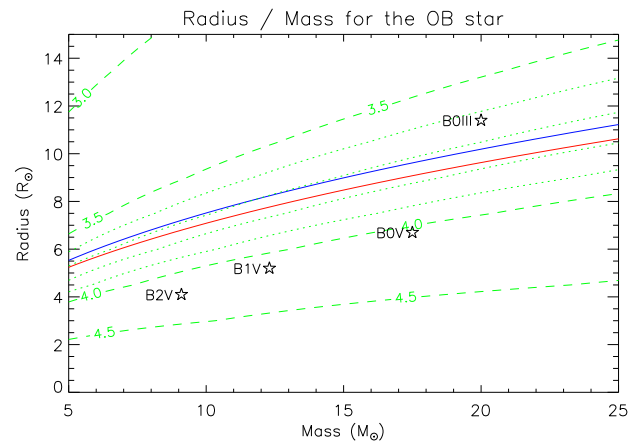} 
  \caption{Mass-Radius relation for the OB star. The solid lines show the Roche lobe radius for the OB star, assuming a 2.4\,\msun\ (upper blue line) and a 1.4\,\msun\ (lower red line) neutron star. The dashed green lines are contours for $\log(g)$ in steps of 0.5\,dex, for which \citet{msc+05} synthetic spectra are available. In the range $3.5 < \log(g) < 4.0$, dotted $\log(g)$ contours are separated by steps of 0.1\,dex. The stars denote the masses and radii of typical isolated B stars.}
  \label{fig:rocherad}
\end{figure}

The Roche lobe radius as a function of B star mass is plotted in \figref{fig:rocherad} for various neutron star masses. For a Roche-filling companion, we infer $3.6 \leq \log(g) \leq 3.8$. For a NS mass in the range 1.4 -- 2.4\,\msun\ and B star mass 8 -- 20\,\msun, the B star radius lies in the range 6 -- 10\,\rsun. Further, the assumed synchronous rotation requires the surface rotational velocity to be in the range $200{\rm\,km\,s}^{-1} \lesssim v_{\rm rot} \lesssim 285{\rm\,km\,s}^{-1}$. These values are consistent with those derived in Section~\ref{subsec:starpar}.

\begin{figure*}[htbp]
  \centering
  \includegraphics[scale=1.0,angle=90,bb=30 30 288 432]{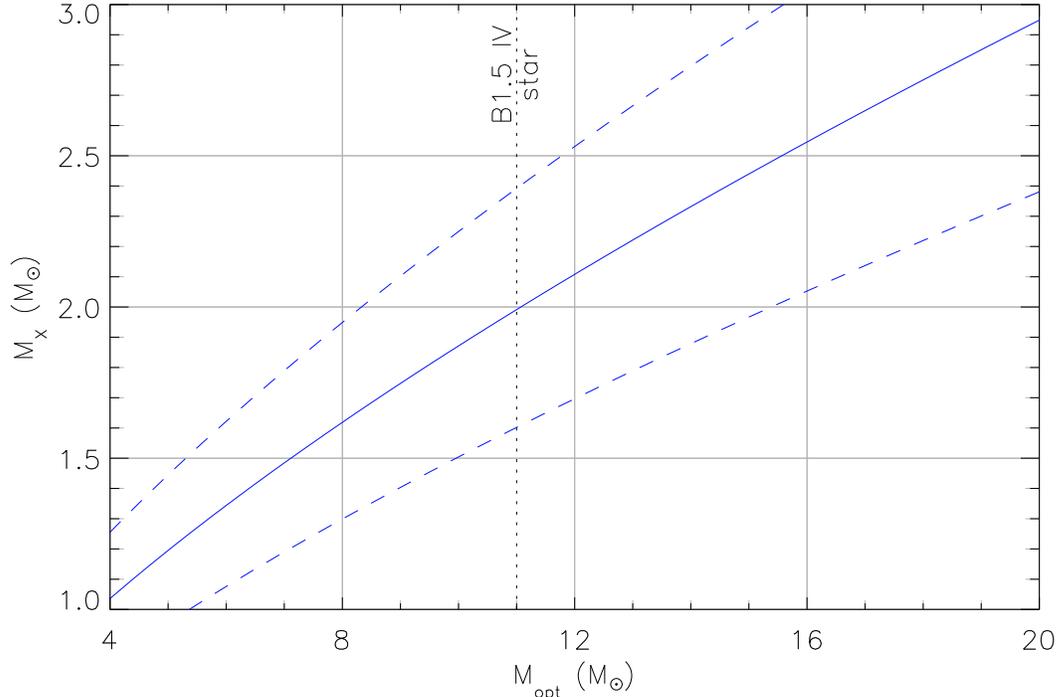}
  \caption{The solid blue line shows the compact object mass ($M_{\rm X}$) as a function of OB star mass ($M_{\rm opt}$). Dashed blue lines are $\pm 1$-$\sigma$ errors. The estimated mass of the B1.5IV primary is 11\,\msun, marked by the vertical dotted line. The corresponding mass of the neutron star is $2.0\pm0.4$\,\msun.}
  \label{fig:massguess}
\end{figure*}

\begin{figure*}[!hbpt]
  \centering
  \includegraphics[scale=0.4]{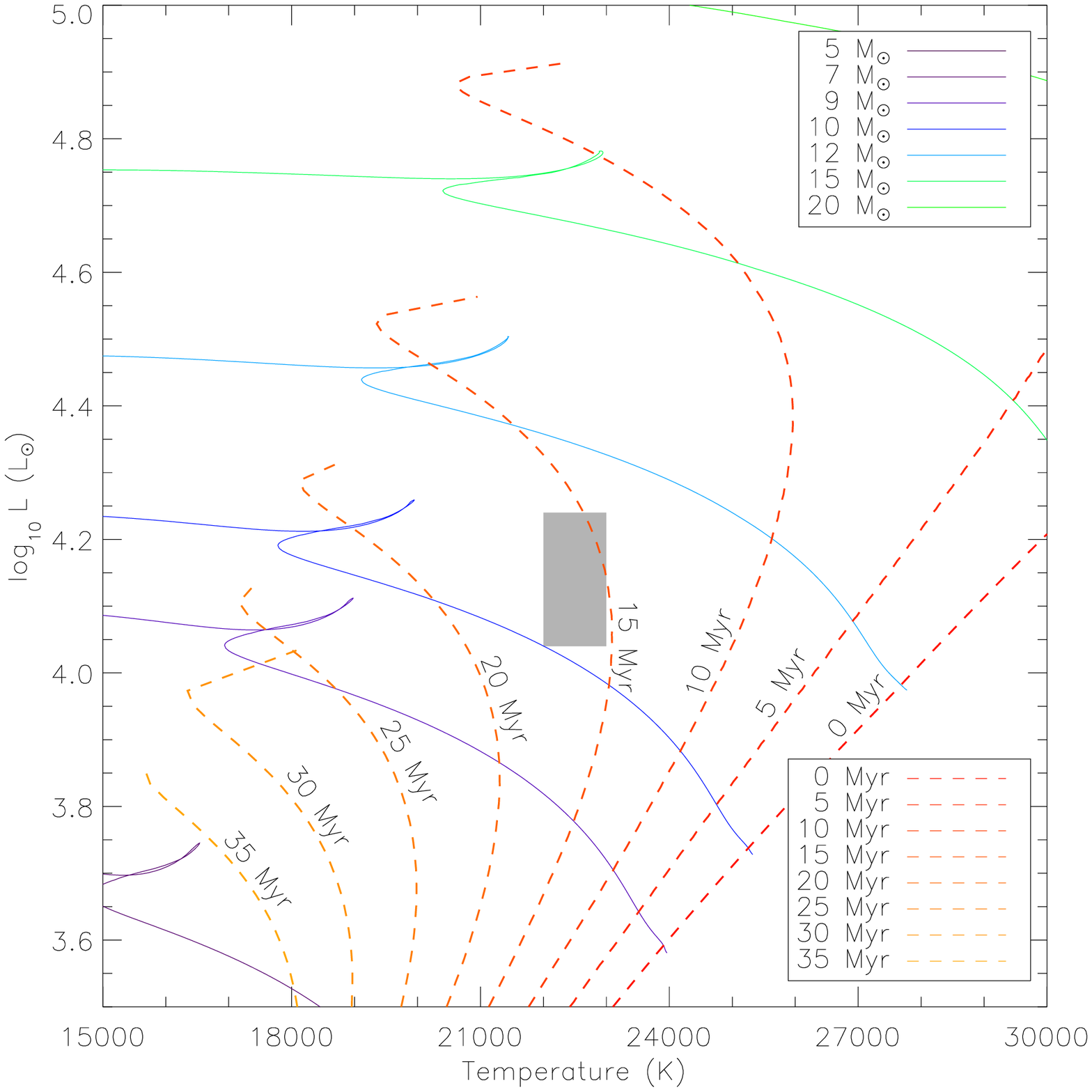}
  \includegraphics[scale=0.4]{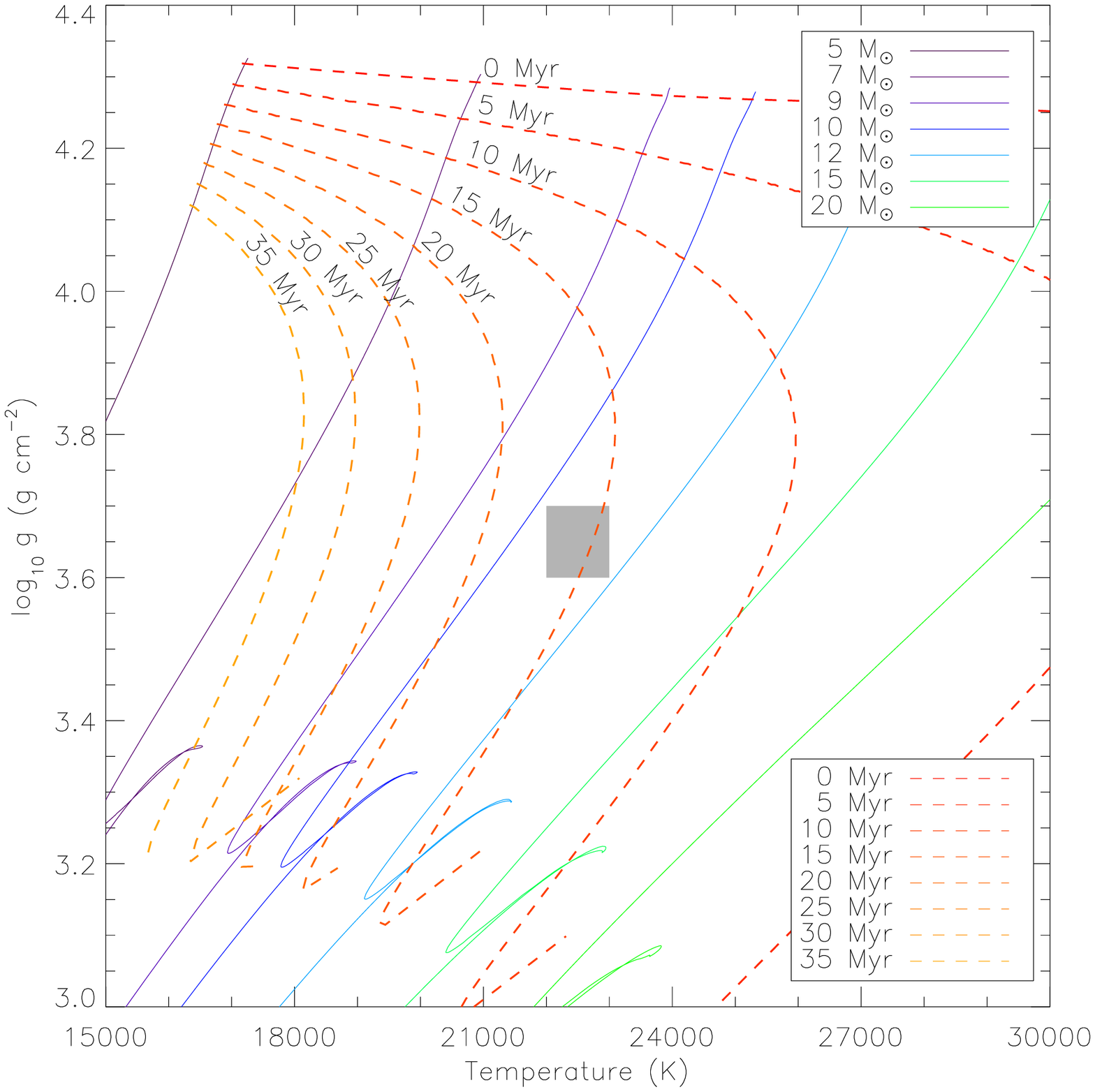}
  \caption{Evolutionary tracks and isochrones for massive stars \citep[adapted from][]{bmc+11}. Left panel: the conventional HR diagram with luminosity and temperature. The primary star has $22000{\rm\,K} \lesssim T \lesssim 23000{\rm\,K}$. We calculate luminosity from the observed $m_V$, and allow a 0.5\,mag offset to calculate the lower limit (see discussion in Section~\ref{subsec:specmass}). This region is shown by a shaded gray box. Right panel: same, but plotted as \logg\ versus $T$. \logg\ for the primary is constrained from Roche lobe arguments (Section \ref{subsec:starpar}). From both panels, we see that the primary is consistent with a 11\,\msun, 15\,Myr object.}
  \label{fig:tracks}
\end{figure*}

Figure~\ref{fig:massguess} plots $M_{\rm X}$ as a function of $M_{\rm opt}$, calculated by solving Equations~\ref{eq:mx}, \ref{eq:sinithetae} and \ref{eq:rla} under known constraints.  If we know $M_{\rm opt}$, we can calculate $M_{\rm X}$. We use the physical properties of the primary (Section~\ref{subsec:starpar}) to estimate the mass of the primary by comparing it with stellar evolutionary models, assuming that binary evolution has not drastically changed the mass-luminosity relation. First, we place the primary on a HR diagram (\figref{fig:tracks}) using models by \citet{bmc+11}. We conservatively allow for a 0.5\,mag error in luminosity. We also plot evolutionary tracks on a \logg\ -- $T$ figure, to utilize the stricter constraints on \logg\ from Roche lobe arguments. From these plots, we see that the primary is approximately a 15\,Myr old, $\sim 11\,$\msun\ star. This is consistent with the typical mass of a B1.5IV star \citep{cox00}.

For $M_{\rm opt} = 11\,$\msun, we calculate $M_{\rm X} = 2.0 \pm
0.4\,$\msun. From evolutionary tracks, we estimate that the
uncertainty in $M_{\rm opt}$ is $\sim 1$\,\msun\
(\figref{fig:tracks}), corresponding to $\Delta M_{\rm X} = 0.12$, much smaller than the uncertainty arising from $\Delta K_{\rm opt}$.
Adding this in quadrature with the uncertainty in the $M_{\rm X}$ --
$M_{\rm opt}$ conversion, we conclude $M_{\rm X} = 2.0 \pm
0.4\,$\msun.


\subsection{Masses from Roche lobe constraints}\label{subsec:rochelims}

\begin{figure*}[!bhtp]
  \centering
  \includegraphics[clip=true,trim=0 0 -10 0,scale=0.55,angle=-90]{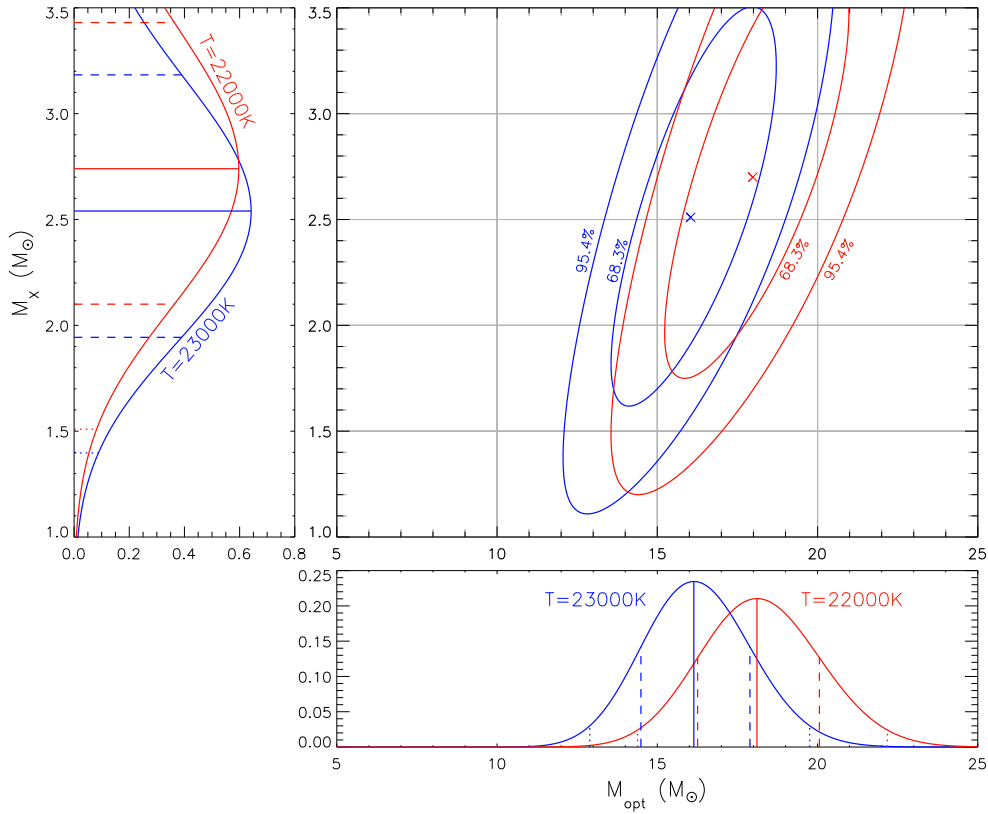} 
  \caption{Probability density plot for neutron star mass ($M_{\rm X}$) as a function of OB star mass ($M_{\rm opt}$). Red contours show the 68.3\% and 95.4\% confidence intervals for masses using the best-fit template  ($\log(g) = 3.5$ and $T = 22000\,$K). The panel on the left shows the PDF for $M_{\rm X}$ marginalized over $M_{\rm opt}$. Similarly, the lower panel shows the PDF for $M_{\rm opt}$, marginalize over $M_{\rm X}$. In these panels, the the solid, dashed and dotted lines show the peak and 68.3\%, 95.4\% confidence intervals respectively. We obtain $M_{\rm X} = 2.7^{+0.7}_{-0.6}$\,\msun, $M_{\rm opt} = 18.1^{+2.0}_{-1.9}$\,\msun. Contours for the the less likely scenario with $\log(g) = 4.0$ and $T = 23000\,$K are shown in blue. In this case, $M_{\rm X} = 2.5\pm0.6$\,\msun, $M_{\rm opt} = 16.1^{+1.8}_{-1.7}$\,\msun.}
  \label{fig:mass}
\end{figure*}

In Section ~\ref{subsec:starpar}, we calculated the radius of the primary from its apparent magnitude, temperature and the distance to M33 (Equation~\ref{eq:rt}). Since the primary is filling its Roche lobe, the stellar radius is equal to the Roche lobe radius ($R_{\rm L}$). This additional constraint can be used in Equations~\ref{eq:mx}~--~\ref{eq:rla} to solve for $M_{\rm X}$ and $M_{\rm opt}$.

We calculate the probability density function (PDF) of component masses as follows. For every pair of assumed masses ($M_{\rm X}$, $M_{\rm opt}$), we use the period $P$ to calculate the semi-major axis $a$. Then we calculate $R_{\rm L}/a$ from \eqnref{eq:rla} and substitute it in Equation~\ref{eq:sinithetae} to calculate $\sin i$. Using $P$, $a$ and $\sin i$ we calculate the expected semi-amplitude of the radial--velocity: 
\begin{equation}
K_2 = \frac{2\pi a \sin i}{P} \frac{M_{\rm X}}{(M_{\rm X} + M_{\rm opt})}
\end{equation}

Next, we calculate the probability for obtaining a certain value of $R_{\rm L}$ and $K_2$, given the measured radius $R$ (Section \ref{subsec:starpar}) and $K_{\rm opt}$ (Section \ref{subsec:orbit}):
\begin{equation}\label{eq:pdf}
  P(R_{\rm L}, K_2) = \exp\left(-\frac{(R_{\rm L} - R)^2}{2\cdot\Delta R^2}\right) \exp\left(-\frac{(K_2 - K_{\rm opt})^2}{2\cdot\Delta K_{\rm opt}^2}\right)
\end{equation}
Here, we are making a simplifying assumption that the Roche volume radius (\eqnref{eq:rla}) is same as the effective radius from photometry (\eqnref{eq:rt}). We convert this PDF to a probability density as a function of $M_{\rm X}$, $M_{\rm opt}$ by multiplying by the 
Jacobian $\partial(R,K_{\rm opt})/\partial(M_{\rm X},M_{\rm opt})$. 

The results are shown in \figref{fig:mass}. Red contours show the 68.3\% and 95.4\% confidence intervals for masses for the best-fit template ($\log(g) = 3.5$ and $T = 22000\,$K). The panel on the left shows the PDF for $M_{\rm X}$ marginalized over $M_{\rm opt}$. Similarly, the lower panel shows the PDF for $M_{\rm opt}$, marginalized over $M_{\rm X}$. In these panels, the the solid, dashed and dotted lines show the peak and 68.3\%, 95.4\% confidence intervals respectively. We obtain $M_{\rm X} = 2.7^{+0.7}_{-0.6}$\,\msun, and $M_{\rm opt} = 18.1^{+2.0}_{-1.9}$\,\msun. For completeness, fits for the less likely scenario with $\log(g) = 4.0$ and $T = 23000\,$K are shown in blue. In this case, $M_{\rm X} = 2.5\pm0.6$\,\msun, $M_{\rm opt} = 16.1^{+1.8}_{-1.7}$\,\msun. 


\begin{deluxetable*}{lr@{\,}lr@{\,}l}
\tablecaption{System parameters for \smc\ and \lmc \label{tab:compare}}
\tablewidth{0pt}
\tablehead{
\colhead{Property} & \multicolumn{2}{c}{\smc} & \multicolumn{2}{c}{\lmc}
}
\startdata
Period ($P$) &  \multicolumn{2}{c}{3.89\,d} & \multicolumn{2}{c}{1.41\,d} \\
P$_{\rm spin}$ &  \multicolumn{2}{c}{0.708\,s} & \multicolumn{2}{c}{13.5\,s} \\
$a_X \sin i$ (lt-s) & 53.4876 & $\pm 0.0004$ & 26.343 & $\pm 0.016$ \\
Eclipse half-angle ($\theta_e$) & 26\degr & --\,30.5\degr & 27\degr & $\pm 2$\degr\\
Mean systemic velocity ($\gamma_{\rm opt}$) & $-191$ & $\pm 6{\rm\,km\,s}^{-1}$ & $306$ & $\pm 10{\rm\,km\,s}^{-1}$ \\
Velocity semi-amplitude ($k_{\rm opt}$) & $20.2$ & $\pm 1.1{\rm\,km\,s}^{-1}$ & $35.1$ & $\pm 1.5{\rm\,km\,s}^{-1}$ \\
\tableline
Companion spectral type & \multicolumn{2}{c}{B0I} & \multicolumn{2}{c}{O8III} \\
Companion $T_{\rm eff}$ &  \multicolumn{2}{c}{29000\,K} & \multicolumn{2}{c}{35000\,K} \\
Distance ($d$) & 60.6 & $\pm 1$\,kpc\tablenotemark{a} & 49.4 & $\pm 1$ \,kpc\tablenotemark{b} \\
Visual magnitude ($m_V$)\tablenotemark{c} & 13.3 & $\pm 0.1$ & 14.0 & $\pm 0.1$ \\
Extinction ($A_V$)\tablenotemark{d} & 0.12 & $\pm 0.01$ & 0.25 & $\pm 0.04$ \\
\tableline
\sidehead{NS mass ($M_{\rm X}$)}
\citet{vkv+07}  & 1.06 & $^{+0.11}_{-0.10}$\,\msun & 1.25 & $^{+0.11}_{-0.10}$\,\msun \\
Our calculation  & 1.32 & $^{+0.16}_{-0.14}$\,\msun & 2.05 & $^{+0.24}_{-0.23}$\,\msun \\
\sidehead{OB star mass ($M_{\rm opt}$)}
\citet{vkv+07} & $\approx$ & 15.7\,\msun & $\approx$ & 14.5\,\msun \\
Our calculation & 21.1 & $^{+3.2}_{-2.9}$\,\msun & 29.1 & $^{+4.8}_{-4.4}$\,\msun
\enddata
\tablecomments{Data from \citet{vkv+07}.}
\tablenotetext{a}{\cite{hhh05}.}
\tablenotetext{b}{\citet{fmb+01}.}
\tablenotetext{c}{Conservative 0.1\,mag errors assumed.}
\tablenotetext{d}{Foreground extinction to galaxy only. Error bars are approximate, the dominant uncertainty has been assigned to $m_V$.}
\end{deluxetable*}

To test the validity of this technique, we apply it to two well--studied targets: \lmc\ and \smc. We find that the basic application of our method is over--estimating the mass (\tabref{tab:compare}). Part of this discrepancy is likely related to equating the Roche volume radius to the effective photometric radius (\eqnref{eq:pdf}). The Roche volume radius (\eqnref{eq:rla}) is the radius of a sphere with the same volume as the Roche lobe of the star. The effective photometric radius (\eqnref{eq:rt}) is the radius of a sphere with the same surface area as the star.  Since an ellipsoid has a larger surface area than a sphere of the same volume, the actual Roche volume radius will be smaller than the effective photometric radius. 
Using a larger radius for the Roche lobe increases the masses of both the components in the binary.

Putting it another way, if we use the Roche volume radius to calculate the brightness of the star, we will get a number lower than the observed brightness. A similar discrepancy is observed by \citet{mmn+12} in massive binaries in the LMC. They find that the absolute magnitude of LMC~172231 calculated using a spherical approximation is 0.45\,mag fainter than observed, while for the triple system [ST92]2-28, the numbers are consistent within errors. Using system parameters derived by \citet{vkv+07}, we find a similar offset of 0.45\,mag for \lmc\ and 0.2\,mag for \smc. If we incorporate this uncertainty by allowing offsets of 0 -- 0.4\,mag, we get $M_{\rm X} = 2.2^{+0.8}_{-0.6}$\,\msun\ and $M_{\rm opt} = 13\pm4$\,\msun. 

Thus while this method has potential, more detailed modeling of the primary is clearly required to accurately infer component masses.

\section{Conclusion}\label{sec:discussion}

From our spectroscopic measurements we find that the donor star in \target\ is a B1.5IV sub-giant with effective temperature $T=22000 - 23000$\,K.  
Assuming a circular orbit, we measure a mean systemic velocity $\gamma_{\rm opt} = -80 \pm 5{\rm\,km\,s}^{-1}$ and velocity semi-amplitude $K_{\rm opt} = 63 \pm 12{\rm\,km\,s}^{-1}$ for the B star. M33 is nearly face-on, with recession velocity of $-179{\rm\,km\,s}^{-1}$ \citep{ddc+91} - so this binary seems to be moving away from the disc at $100{\rm\,km\,s}^{-1}$.

Using the physical properties of the B star determined from our optical spectroscopy we find a
mass for the donor of $M_{\rm opt} = 11\pm1$\,\msun.    This mass is based on stellar evolution models, and will be
reasonably accurate so long as binary evolution has not significantly altered the mass-luminosity relation.  However,
it is difficult to test this assumption based on any available observations.   Using this spectroscopic mass,
we calculate the mass of the compact object, $M_{\rm X} = 2.0 \pm0.4$\,\msun. 
This is higher than the canonical 1.4\,\msun\ for neutron stars, but comparable to masses of other neutron stars in X-ray binaries such as the
HMXB Vela~X$-$1 \citep[$1.88\pm0.13$\,\msun;][]{bkv+01,qna+03} or the Low Mass X-ray binaries Cyg~X$-$2 \citep[$1.71\pm0.21$\,\msun;][]{cgi+10} and 4U~1822$-$371 \citep[$1.96\pm0.35$\,\msun;][]{mcm05}.    Since no pulsations have been detected we have only
indirect evidence, based on the X-ray spectrum, that the compact object is a neutron star.   However, the mass we derive here
is smaller than would be expected for a black hole.

Based on the stable X-ray flux, we infer that the 
donor is transferring mass to the neutron star by Roche lobe overflow.   By equating the Roche lobe radius to physical radius of  $R = 9.1\pm0.3\,R_\odot$,
derived from the known distance to M33, combined with the stellar luminosity and temperature, we derive an additional orbital constraint. 
"From a first pass calculation with a spherical approximation for the
shape of the primary, we find substantially larger masses. However,
applying this technique to the well-studied binaries LMC~X-4 and SMC~X-1, 
both of which have measured component masses, we find it
consistently overestimates the compact object mass. This is likely
because the Roche surface is not spherical but elongated, which is not
taken into account in our estimate. Attempting to account for that, we
infer $M_{\rm X}= 2.7^{+0.7}_{-0.6}$\,\msun\ and $M_{\rm opt}=
18.1\pm2.0$\,\msun.  Future efforts to more accurately model the system
geometry will improve the accuracy of this technique, which is
applicable generally to Roche lobe overflow systems with known
distances.

\section*{Acknowledgments}
We thank Brian Grefenstette for helping interpret XMM data.

Some of the data presented herein were obtained at the W.M. Keck Observatory, which is operated as a scientific partnership among the California Institute of Technology, the University of California and the National Aeronautics and Space Administration. The Observatory was made possible by the generous financial support of the W.M. Keck Foundation.

This research has made use of the NASA/IPAC Extragalactic Database (NED) which is operated by the Jet Propulsion Laboratory, California Institute of Technology, under contract with the National Aeronautics and Space Administration. This research has made use of NASA's Astrophysics Data System Bibliographic Services. This research used the facilities of the Canadian Astronomy Data Centre operated by the National Research Council of Canada with the support of the Canadian Space Agency.


\end{document}